\documentclass[letter,twocolumn]{jpsj3}

\title{Electronic Ferroelectricity in a Dimer Mott Insulator}
\author{Makoto~Naka, and Sumio~Ishihara~\thanks{E-mail address:ishihara@cmpt.phys.tohoku.ac.jp} }
\inst{Department of Physics, Tohoku University, Sendai 980-8578, Japan}
\abst{
Motivated from recent experiments in $\kappa$-(BEDT-TTF)$_2$Cu$_2$(CN)$_3$, dielectric and magnetic properties in a dimer-Mott insulator are studied. 
We derive the effective Hamiltonian where the number of electron per dimer is taken to be one. 
Charge and magnetic structures in finite temperature are obtained by using the mean-field  and Monte-Carlo methods. 
Magnetic and ferroelectric phases are exclusive with each other. 
Dielectric fluctuation is remarkable near a phase boundary between dimer-Mott phase and ferroelectric  phase.  Implications of the present results for $\kappa$-(BEDT-TTF)$_2$Cu$_2$(CN)$_3$ are discussed. 
}
\kword{Ferroelectricity, Frustration, Charge Order}

\begin{document}
\maketitle

Novel dielectric and magneto-dielectric phenomena are one of the recent central issues in solid state physics. 
Beyond a conventional picture based on classical dipole moments, 
ferroelectricity on which electronic contribution plays crucial roles is termed electronic ferroelectricity.~\cite{brink,ishihara} 
Recently discovered multiferroics are known as phenomena where ferroelectricity is driven by a spin ordering. 
In a Mott insulator with frustrated exchange interactions, 
an electric polarization is induced by the exchange striction effects under a non-collinear spin structure. 

There is another class of electronic ferroelectricity; charge-order driven ferroelectricity where electric polarization is caused by an electronic charge order (CO) without inversion symmetry. 
This class of ferroelectricity is observed in transition-metal oxides and charge-transfer type organic salts, 
for exmple LuFe$_2$O$_4$,~\cite{ikeda,nagano,naka} (TMTTF)$_2$X (X: a monovalent cation),~\cite{monceau,yoshioka,otsuka} 
and $\alpha$-(BEDT-TTF)$_2$I$_3$.~\cite{yamamoto} 
A large magneto-dielectric coupling and fast polarization switching are expected in charge-driven ferroelectricity, since the electric polarization is governed by electrons. 

Dielectric anomaly recently discovered in a quasi-two dimensional organic salt $\kappa$-(BEDT-TTF)$_2$Cu$_2$(CN)$_3$ suggests a possibility of electronic ferroelectricty in this compound. 
The crystal structure consists of an alternate stacking of the BEDT-TTF donor layers and the Cu$_2$(CN)$_3$ acceptor layers. In a quasi-two-dimensional BEDT-TTF layer, pairs of dimerized molecules locate in an almost equilateral triangular lattice. 
When two dimerized molecules are considered as a unit, 
since average hole number per dimer is one, this material is identified as a Mott insulator. 
One noticeable property observed experimentally is a low temperature spin state; no evidences of a long-range magnetic order down to 32mK.~\cite{yamashita,shimizu} 
A possibility of quantum spin-liquid states is proposed. 
In the recent experiments~\cite{abel}, temperature dependence of the dielectric constant has a broad maximum around 25K and shows relaxor-like dielectric relaxation. 
Some anomalies are also seen in the lattice expansion coefficient and specific heat around 6K.~\cite{yamashita2,manna} 
These data promote us to reexamine electronic structure in dimer Mott (DM) insulators. 

In this Letter, motivated from the recent experimental results in $\kappa$-(BEDT-TTF)$_2$Cu$_2$(CN)$_3$, we study dielectric and magnetic properties in a DM insulating system. From a Hubbard-type Hamiltonian, we derive the effective Hamiltonian where the number of electron per dimer is one. 
By using the mean-field (MF) approximation and the classical Monte-Carlo (MC) simulation, 
we examine spin and charge structures in finite temperature. 
It is shown that the ferroelectric and magnetic phases are exclusive with each other. 
A reentrant feature of the DM phase enhances the dielectric fluctuation near the CO phase. 
Implications of the present results for $\kappa$-(BEDT-TTF)$_2$Cu$_2$(CN)$_3$ 
are discussed. 

We start from the model Hamiltonian to describe the electronic structure in a DM insulator. 
A moleculer-dimer is regarded as a unit and is allocated at each site of a two-dimensional triangular lattice. 
An average electron number per dimer is assumed to be one. 
The Hamiltonian consists of the two terms as 
\begin{eqnarray}
{\cal H}_0={\cal H}_{\rm intra}+{\cal H}_{\rm inter} . 
\label{eq:h0}
\end{eqnarray}
The first term is for the intra-dimer part given  by 
\begin{align}
{\cal H}_{\rm intra}
&= \varepsilon\sum_{i \mu s} c_{i \mu s}^\dagger c_{i \mu s}^{}
-t_0 \sum_{i s} \left ( c_{i a s}^\dagger c_{i b s}^{}+ H.c. \right ) 
\nonumber \\
&+ U_0 \sum_{i \mu } n_{i \mu \uparrow} n_{i \mu \downarrow}
+V_0 \sum_{i } n_{i a} n_{i b} ,
\label{eq:hintra}
\end{align}
where two molecules are identified by a subscript $\mu(=a,b)$. 
We introduce the electron annihilation operator $c_{i \mu s}$ for molecule $\mu$, spin $s(=\uparrow, \downarrow)$ at site $i$, and the number operator 
$n_{i \mu}=\sum_{s} n_{i \mu s}=\sum_{s} c_{i \mu s}^\dagger c_{i \mu s}$. 
We consider a level energy $\varepsilon$, the inter-molecule electron transfer $t_0(>0)$ in a dimer, the intra-molecule Coulomb interaction $U_0$ and the inter-molecule Coulomb interaction $V_0$ in a dimer.  
In addition, the inter-dimer part in the second term of Eq.~(\ref{eq:h0}) is given by 
\begin{align}
{\cal H}_{\rm inter}
&=-\sum_{\langle ij \rangle \mu \mu' s}
t_{ij}^{\mu \mu'} 
\left ( c_{i \mu s}^\dagger c_{j \mu' s}+H.c. \right )
\nonumber \\
&+ \sum_{\langle ij \rangle \mu \mu'} V_{i j}^{\mu \mu'} n_{i \mu } n_{j \mu'}, 
\label{eq:hinter} 
\end{align}
where $t_{ij}^{\mu \mu'}$ and $V_{ij}^{\mu \mu'}$ are the electron transfer and the Coulomb interaction between an electron in a molecule $\mu$ at site $i$ and that in a molecule $\mu'$ at site $j$, respectively. The first and second terms in ${\cal H}_{\rm inter}$ are denoted by ${\cal H}_t$ and ${\cal H}_V$, respectively. 

We briefly introduce the electronic structure in an isolated dimer. 
In the case where one electron occupies a dimer, 
the bonding and anti-bonding states are given by 
$|\beta_s \rangle =(|a_s \rangle + | b_s \rangle)/\sqrt{2}$ 
and 
$|\alpha_s \rangle =(|a_s \rangle - | b_s \rangle)/\sqrt{2}$ 
with energies  
$E_\beta = \varepsilon - t_0$ and 
$E_\alpha = \varepsilon + t_0$, respectively. 
In these bases, we introduce the electron operator 
$\hat c_{i \gamma s}$ for $\gamma=(\alpha, \beta)$ 
and the electron transfer integral ${\hat t}^{\gamma \gamma'}_{ij}$ 
between the NN molecular orbitals $\gamma$ and $\gamma'$. 
These are obtained by the unitary transformation 
from $c_{i \mu s}$ and $t_{ij}^{\mu \mu'}$. 
Two-electron states in a dimer are the following six states: 
the spin-triplet states 
$\{ |T_{\uparrow} \rangle, | T_\downarrow \rangle , | T_0 \rangle \}  = \{ 
|\alpha_\uparrow \beta_\uparrow \rangle,
|\alpha_\downarrow \beta_\downarrow \rangle, 
(|\alpha_\uparrow \beta_\downarrow \rangle+|\alpha_\downarrow \beta_\uparrow \rangle)/\sqrt{2}   \} $  
with the energy $E_T=2\varepsilon+V_0$, 
the spin-singlet state 
$|S \rangle =(|\alpha_\uparrow \beta_\downarrow \rangle -|\alpha_\downarrow \beta_\uparrow \rangle) /\sqrt{2} $
with $E_{S}=2\varepsilon+U_0$, 
and 
the doubly-occupied states 
$|D_+ \rangle=C_1| \alpha_\uparrow \alpha_\downarrow \rangle + C_2|\beta_\uparrow \beta_\downarrow \rangle$ 
and 
$|D_- \rangle=C_2| \alpha_\uparrow \alpha_\downarrow \rangle - C_1|\beta_\uparrow \beta_\downarrow \rangle$ 
with $E_{D\pm}=(4\varepsilon+U_0+V_0 \pm \sqrt{(U_0-V_0)^2+16t_0^2} )/2$ and 
coefficients 
$C_2/C_1=(U_0-V_0)/[2E_{D+}-4(\varepsilon-t_0)-(U_0-V_0)]$. 
The lowest eigen state is $|D_- \rangle $. 
The effective Coulomb interaction in the lowest eigen state is $U_{eff} \equiv E_{D_-}-2E_\beta \sim V_0+2t_0$ in the limit of $U_0, V_0 >> t_0$. 

\begin{figure}[t]
\begin{center}
\includegraphics[width=0.7\columnwidth,clip]{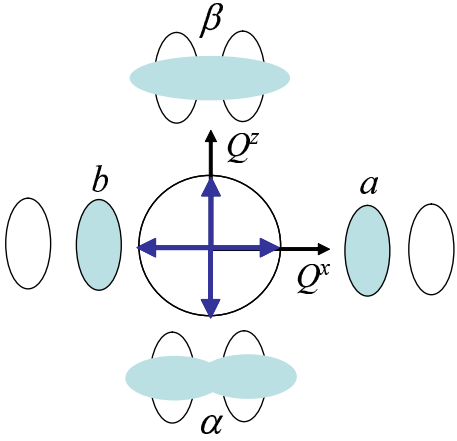}
\end{center}
\caption{(Color online) 
Pseudo-spin directions in the $Q^x-Q^z$ plane and electronic structures in a dimer. 
}
\label{fig:ps}
\end{figure}
%
From the Hamiltonian in Eq.~(\ref{eq:h0}), 
we derive an effective model in the subspace where each dimer is occupied by one electron. 
Charge structure in each dimer is represented by the pseudo-spin (PS) operator 
with amplitude of 1/2 defined by 
$
{\bf Q}_{i}=\frac{1}{2} \sum_{s \mu \mu'} {\hat c}_{i \mu s}^\dagger {\bf \sigma}_{\mu \mu'} {\hat c}_{i \mu' s}^{} 
$
with the Pauli matrices $\bf \sigma$. 
The eigen states for $Q^x_i$ with the eigen values of $1/2$ and $-1/2$ are the charge polarized states where an electron occupies the $a$ and $b$ molecules, respectively, 
and those for $Q^z_i$ with $1/2$ and $-1/2$ correspond to 
the covalent states where an electron occupies the $\beta$ and $\alpha$ orbitals, respectively (see Fig.~\ref{fig:ps}). 

Here we derive the effective Hamiltonian. 
We assume $U_0, V_0 >> t_{ij}^{\mu \mu'}, V_{ij}^{\mu \mu'}$, 
and treat the inter-dimer term ${\cal H}_{\rm inter}$ in Eq.~(\ref{eq:h0}) as the perturbed term. The effective Hamiltonian up to the orders of $O({\cal H}_t^2)$ and $O({\cal H}_V^1)$ is given by 
$
{\cal H}={\widetilde {\cal H}}_{\rm intra}+{\widetilde {\cal H}}_{V}+{\cal H}_J . 
$
The first and second terms correspond to ${\cal H}_{\rm intra}$ in Eq.~(\ref{eq:hintra}) and the second term of ${\cal H}_{\rm inter}$ in Eq.~(\ref{eq:hinter}), respectively, where the doubly occupied states in a dimer are prohibited. 
By using the PS operator, these terms are given by 
\begin{align}
{\widetilde {\cal H}}_{\rm intra}+{\widetilde {\cal H}}_{V}=-
2t_0\sum_{i} Q^z_{i}+\sum_{\langle ij \rangle} W_{ij} Q_i^x Q_j^x, 
\label{eq:tising}
\end{align}
where $W_{ij}(=V_{ij}^{aa}+V_{ij}^{bb}-V_{ij}^{ab}-V_{ij}^{ba})$ 
is the effective inter-dimer Coulomb interaction. 
This part is nothing but the transverse-field Ising model. 

The third term in ${\cal H}$ is the exchange-interaction term derived by the second-order perturbation with respect to ${\cal H}_{t}$ in Eq.~(\ref{eq:hinter}).
This is given by a sum of the terms classified by the two-electron states in a dimer 
denoted by $m$ as 
$
{\cal H}_J=\sum_{m} {\cal H}_J^{(m)} ,
$
where a suffix $m=\{ T_\uparrow, T_\downarrow, T_0, S, D_+, D_- \}$. 
A dominant term in ${\cal H}_J$ is ${\cal H}_J^{(D_-)}$ which has the lowest intermediate state energy. This is explicitly given by 
\begin{align}
{\cal H}_{J}^{(D-)}=-\sum_{\langle ij \rangle} 
\left ( \frac{1}{4} - {\bf S}_i \cdot {\bf S}_j \right ) 
h_{ij}^{(D-)} , 
\label{eq:Hd-}
\end{align}
with 
\begin{align}
h_{ij}^{(D-)}
&=\sum_{\gamma_1, \gamma_2=(\alpha, \beta)} 
J^{\gamma_1 \gamma_2}_{ij} n_{i \gamma_1} n_{j \gamma_2}
+\sum_{\nu_1, \nu_2=(+, -)}
J^{\nu_1 \nu_2}_{ij} Q_i^{\nu_1} Q_j^{\nu_2}
\nonumber \\
&+\sum_{\gamma=(\alpha, \beta)} 
\left ( J^{x \gamma}_{ij} Q_i^x n_{j \gamma} +J^{\gamma x}_{ij} n_{i \gamma}Q_{j}^x  \right ) . 
\label{eq:hd-}
\end{align}
We define $Q^{\pm}_i=Q_i^x \pm iQ_i^y$ and $n_{i \alpha(\beta)}=1/2-(+) Q_i^z$. 
Expressions of the exchange constants are given in Ref.~\cite{exch}. 
Other terms in ${\cal H}_J$  
are represented by similar forms with Eqs.~(\ref{eq:Hd-}) and (\ref{eq:hd-}). 
Spin and charge degrees are coupled with each other in the Hamiltonian, although the SU(2) symmetry is preserved only in the spin sector. 
This type of the Hamiltonian is similar to the so-called Kugel-Khomskii model~\cite{kugel}  
in an orbital degenerated Mott insulator, and was also proposed in study of the electronic state in $\alpha$-NaV$_2$O$_5$.~\cite{thalmeier,mostovoy,sa} 

This Hamiltonian is analyzed by the MF approximation and the classical MC method. 
In the MF calculations, 
triangular lattices of 12 unit cells along $\langle 110 \rangle$ direction with the periodic boundary condition are used. 
We adopt the following 15 MF's, $\langle S^\mu \rangle$, $\langle Q^\mu \rangle$ and $\langle S^\mu Q^\nu \rangle$ where $(\mu, \nu)=(x,y,z)$, in each $ \langle 1{\bar 1}0 \rangle$ line.  
The multi-canonical MC simulations are performed in finite-size clusters of $L$ sites ($L\le 96$) with the periodic-boundary condition. 
We use 10$^7$ MC steps to obtain histograms and $2\times 10^7$ steps for measurements. 

\begin{figure}[t]
\begin{center}
\includegraphics[width=1.1\columnwidth,clip]{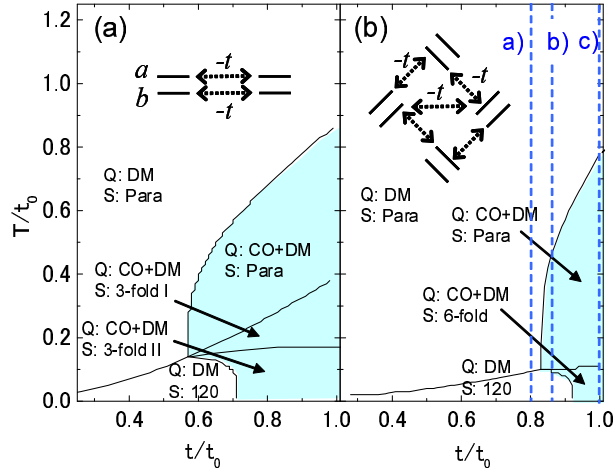}
\end{center}
\caption{(Color online) 
Finite-temperature phase diagrams obtained by MF method. 
(a) and (b) are for case I and case II, respectively (see text). 
Symbols Q:DM and Q:CO+DM represent the DM phase, and the coexistence phase of CO and DM states, respectively, and S:Para, S:120, S:3-fold I, S:3-fold II, and S:6-fold represent the paramagnetic phase, the 120$^\circ$ structure phase, the collinear 3-fold spin ordered phase, the coplanar 3-fold spin ordered phase, and the coplanar 6-fold spin order phase, respectively. 
We chose $W/t_0=1$ in (a), and $U_0/t_0=6$, $V_0/t_0=4.5$, $W_1/t_0=-1$, and $W_2/t_0=0.07$ in (b) where $W_1$ and $W_2$ are the NN inter-dimer Coulomb interactions along 
$\langle 100 \rangle$ and $\langle 110 \rangle$, respectively. 
The insets are schematic pictures of the inter-dimer transfer integrals. 
Vertical broken lines in (b) represent cases a), b), and c) (see text).  
}
\label{fig:phase}
\end{figure}
%
Finite-temperature phase diagrams obtained by the MF method are shown in Fig.~\ref{fig:phase}. As for $W_{ij}$ in Eq.~(\ref{eq:tising}) and $t_{ij}^{\mu \nu}$ in Eq.~(\ref{eq:hinter}), we consider the following two cases in order to identify general and specific features in the results: 
case I for a simple model, and case II for $\kappa$-(BEDT-TTF)$_2$Cu$_2$(CN)$_3$. 
In case I, we chose $t_{ij}^{\mu \nu}=\delta_{\mu \nu} t$ $(t>0)$ and $W_{ij}=W(>0)$. 
In case II, the three dominant $t_{ij}^{\mu \nu}$'s estimated in $\kappa$-(BEDT-TTF)$_2$Cu$_2$(CN)$_3$ are represented by a parameter $t$ as shown in the inset of Fig.~\ref{fig:phase}(b), 
and $W_{ij}$ are considered based on the crystal structural data.~\cite{mori,seo,hotta} 
Similar phase diagrams are obtained in the two cases shown in Figs.~\ref{fig:phase}(a) and 2(b) except for detailed spin structures. 
The CO phases realize in the region of large $t$. This phase is caused by ${\widetilde {\cal H}}_V$ and the term proportional to $Q_i^xQ_j^x$ in ${\cal H}_J$. 
In low temperatures, magnetic phases are stabilized by the exchange interactions.  
One noticeable point is that when the magnetic phase appears, the CO phase is suppressed [see around $t/t_0=0.6$ in Fig.~\ref{fig:phase}(a) and around $t/t_0=0.8$ in Fig.~\ref{fig:phase}(b)]. 
As a result, a reentrant feature is observed in a DM phase. 
The reverse is also seen; 
as shown around $T/t_0=0.1$ and $t/t_0>0.9$ in Fig.~\ref{fig:phase}(b), 
a slope of the magnetic transition temperature versus $t$ curve decreases with increasing $t$, when the CO sets in. That is, the charge and magnetic phases are exclusive with each other. 
This is caused by a competition between the terms proportional to $Q_i^xQ_j^x$ 
and $Q_i^xQ_j^x{\bf S}_i \cdot {\bf S}_j$ in ${\cal H}_J$; 
the former favors a ferro-type configuration of $Q^x$, but the latter favors an antiferro-type one under the antiferromagnetic spin correlation $\langle {\bf S}_i \cdot {\bf S}_j \rangle<0$. 

\begin{figure}[t]
\begin{center}
\includegraphics[width=\columnwidth,clip]{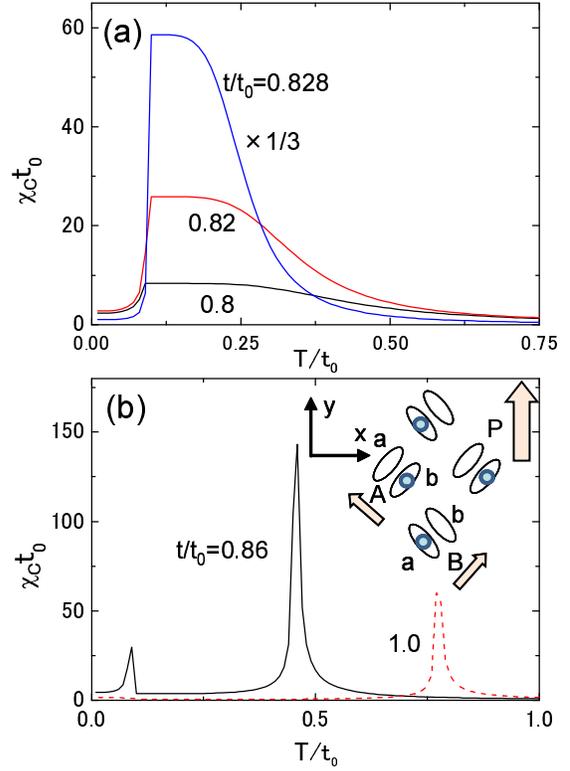}
\end{center}
\caption{(Color online) 
(a) Temperature dependences of the charge susceptibility $\chi_{C}$ along $y$ in case a), 
and (b) those in cases b) and c) obtained by the MF method (see text). 
The inset of (b) is a schematic CO pattern in a triangular lattice. 
Ellipses and circles represent molecules and electrons, respectively. 
}
\label{fig:suscept}
\end{figure}
Let us focus on case II in more detail. 
Temperature dependences of the charge susceptibility are presented in Fig.~\ref{fig:suscept} for several values of $t/t_0$. We define the charge susceptibility along the $y$ direction as  
$\chi_C=(TL)^{-1} (\sum_{i \in A} - \sum_{i \in B} ) (\sum_{j \in A} - \sum_{j \in B} ) $ $[\langle Q_i^x Q_j^x \rangle-\langle Q_i^x \rangle \langle Q_j^x \rangle ]$ 
where two sublattices are identified by $A$ and $B$ [see the inset of Fig.~\ref{fig:suscept}(b)]. We pay our attention to a region of $t/t_0=0.8-1$. 
In Fig.~\ref{fig:phase}(b), 
there are three types of sequential phase changes with decreasing $T$; 
a) (Q:DM, S:Para)$\rightarrow$ (Q:DM, S:120), 
b) (Q:DM, S:Para)$\rightarrow$ (Q:CO+DM, S:Para) $\rightarrow$ (Q:DM, S:120), 
and 
c) (Q:DM, S:Para)$\rightarrow$ (Q:CO+DM, S:Para) $\rightarrow$ (Q:CO+DM, S:6-fold) where abbreviations are defined in the caption of Fig.~\ref{fig:phase}.
In the case a), although the CO phase does not realize, $\chi_C$ increases in the paramagnetic phase and suddenly reduces in the low temperature magnetic phase. This enhancement of $\chi_C$ is remarkable at a vicinity of the phase boundary between the CO and DM phases (see a curve for $t/t_0=0.828$ in Fig.~\ref{fig:suscept}).
In the cases b) and c), the CO phase realizes and $\chi_C$ diverges at the phase boundary. 
Obtained CO pattern is schematically shown in the inset of Fig.~\ref{fig:suscept}(b). 
This is a ferri-electric structure and is similar to the one proposed in Ref.~\cite{abel}. 
Within a chain along the $x$ axis, the dipole moments align uniformly. 
Directions of the dipole moments in sublattices $A$ and $B$ are almost perpendicular with each other.  
As a result, the electric polarization appears along the $y$ axis. 
This ordered pattern is energetically favorable for the inter-site Coulomb interaction $W_{ij}$ and the term which is proportional to $Q_i^x Q_j^x$ in ${\cal H}_J$.

\begin{figure}[t]
\begin{center}
\includegraphics[width=\columnwidth,clip]{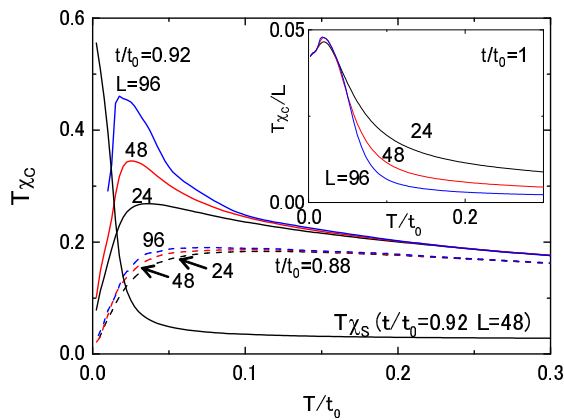}
\end{center}
\caption{(Color online) 
Temperature dependences of the charge susceptibility $T\chi_C$ obtained by the MC simulation in the case a). 
The inset shows the susceptibilities devided by $L$ in the case c). 
The spin susceptibility $T \chi_S$ is also plotted. 
}
\label{fig:mc}
\end{figure}
%
We also examine the dielectric and magnetic structures in case II by the MC method. Temperature dependences of the charge susceptibilities $T\chi_C$ are presented in Fig.~\ref{fig:mc} for several $t/t_0$. 
It is shown that with increasing $L$, $T\chi_C$'s in $t/t_0=0.88$ and 0.92 tend to be constants. 
On the other hand, the susceptibility divided by the system size, $T\chi_C/L$, in $t/t_0=1$ is almost independent of $L$ for $T/t_0<0.05$. 
Therefore, the CO phase realizes in $T/t_0<0.05$ in $t/t_0=1$, and others are the DM phase. 
The results in $t/t_0=0.88$ and 0.92, and the ones in $t/t_0=1$ correspond to the case a) and the case c), respectively, although the corresponding values of $t/t_0$ 
and the ordering temperatures are different from those in the MF calculations. 
With decreasing $T$ in $t/t_0=0.92$, $T\chi_C$ increases and suddenly goes down around $T/t_0=0.03$ where the spin susceptibility 
$\chi_S[\equiv (TL)^{-1} \sum_{i j} e^{i {\bf q} \cdot {\bf r}_{ij}}
\langle {\bf S}_i \cdot {\bf S}_j \rangle]$ 
at ${\bf q}=(1/3,1/3)$, corresponding to the 120$^\circ$ structure, develops. 
The results are consistent qualitatively with the MF calculation results.

Finally, we discuss implications of the present results for the dielectric and magnetic properties in $\kappa$-(BEDT-TTF)$_2$Cu$_2$(CN)$_3$. 
We at first speculate that the transition temperature for the 120$^\circ$ spin structure seen in Fig.~\ref{fig:phase} (b) is interpreted to a temperature where the AFM spin correlation is developed in this compound. 
This is reasonable since this spin structure appears in a DM phase where spin frustration survives and the classical ordering temperature is expected to be largely reduced. On the other hand, calculated spin orders in the CO phase and their transition temperatures are substantial, because the anisotropic exchange interactions in the CO pattern shown in Fig.~\ref{fig:suscept}(b) release spin frustration.  

Based on the present study, we propose two possible scenarios which are relevant to $\kappa$-(BEDT-TTF)$_2$Cu$_2$(CN)$_3$.
In the first scenario, the dipole moments freeze in low temperatures. 
This corresponds to the sequential phase change of the case c) introduced previously; 
(Q:DM, S:Para)$\rightarrow$ (Q:CO+DM, S:Para) $\rightarrow$ (Q:CO+DM, S:6-fold). 
A relaxor-like dielectric dispersion observed experimentally requires 
a random freezing or a dipolar glass state below temperature where the dielectric constant takes a peak. This may be consistent with the experimental results suggesting inhomogeneous magnetic moment in low temperatures~\cite{shimizu2,kawamoto}. 
We propose from the present calculations that the characteristic magnetic temperature and the spin correlation functions in the CO phase are smaller than those in the DM phase, because the CO and magnetic phases are exclusive with each other as shown previously. 
One problem in this scenario is that the random/uniform polarization freezing makes the exchange interaction strongly anisotropy and releases spin frustration. 
This may be unfavorable for the experimental fact that neither a long-range magnetic order nor a spin glass transition appears down to 32mK.~\cite{yamashita,shimizu}

Another possible scenario is that the dipole moments do not freeze in low temperatures, and the dielectric anomaly experimentally observed is caused by charge fluctuation at vicinity of the phase boundary. 
This corresponds to the sequential phase change of the case a); 
(Q:DM, S:Para)$\rightarrow$ (Q:DM, S:120). 
Because of the reentrant feature of the DM phase shown in Fig.~\ref{fig:phase}(b), the system approaches to the phase boundary at certain temperatures, and the charge susceptibility is enhanced as shown in Fig.~\ref{fig:suscept}(a) and Fig.~\ref{fig:mc}. 
With decreasing temperature furthermore, development of the spin correlation stabilizes the DM phase and suppresses the dielectric fluctuation. 
In this scenario, spin frustration is alive in the low temperature DM phase. 
Further examinations 
will provide a clue to unveil the microscopic pictures in the low temperature magnetic and dielectric structures.

The authors would like to thank T.~Sasaki, I.~Terasaki, S.~Iwai, T.~Watanabe, H.~Seo and J. Nasu for valuable discussions. This work was supported by JSPS KAKENHI, TOKUTEI from MEXT, CREST, and Grand Challenges in Next-Generation Integrated Nanoscience. 
MN is supported by the global COE program of MEXT Japan.


\begin{thebibliography}{99}
\bibitem{exch}
The exchange constants in Eq.~(\ref{eq:hd-}) are  
$J^{\gamma \gamma}_{ij}=4 {\hat t}^{\gamma \gamma 2}_{ij} D_{\bar \gamma}^2 \Delta_{\gamma \gamma}^{-1}$, 
$J^{\gamma {\bar \gamma}}_{ij}=2 {\hat t}^{\gamma {\bar \gamma} 2}_{ij} \Delta_{\gamma {\bar \gamma}}^{-1}$, 
$J^{\nu \nu}_{ij}=-2{\hat t}^{\alpha \alpha}_{ij} {\hat t}^{\beta \beta}_{ij} 
C_1 C_2 (\Delta_{\alpha \alpha}^{-1}+\Delta_{\beta \beta}^{-1})$, 
$J^{\nu {\bar \nu}}_{ij}=-4{\hat t}^{\alpha \beta}_{ij} {\hat t}^{\beta \alpha}_{ij} 
C_1 C_2 \Delta_{\alpha \beta}^{-1}$, 
$J^{x \gamma}_{ij}=2{\hat t}^{\gamma \gamma}_{ij} {\hat t}^{{\bar \gamma} \gamma}_{ij} 
D_{{\bar \gamma}} (D_{\bar \gamma}-D_{\gamma}) (\Delta_{\gamma \gamma}^{-1}+\Delta_{\alpha \beta}^{-1})$
and 
$J^{\gamma x}_{ij}=2{\hat t}^{\gamma \gamma}_{ij} {\hat t}^{\gamma {\bar \gamma}}_{ij} 
D_{{\bar \gamma}} (D_{\bar \gamma}-D_{\gamma}) (\Delta_{\gamma \gamma}^{-1}+\Delta_{\alpha \beta}^{-1})$
 with the energy difference 
$\Delta_{\gamma_1 \gamma_2}=E_{D_-}-E_{\gamma_1}-E_{\gamma_2} $. 
We define ${\bar \gamma}=(\alpha, \beta)$ for $\gamma=(\beta, \alpha)$, 
${\bar \mu}=(+,-)$ for $\mu=(-,+)$, and $D_\gamma=(C_1, C_2)$ for $\gamma=(\alpha, \beta)$ 
where $C_1$ and $C_2$ are the coefficients in the wave function $|D_- \rangle$.
\end{thebibliography}
\end{document}